# An Empirical Study of Irregular AG Block Turbo Codes over Fading Channels

Omar A. Alzubi

Al-Balqa Applied University

Salt, Jordan

o.zubi@bau.edu.jo

**Abstract:** This paper will present the design, construction, and implementation of Algebraic-Geometric Irregular Block Turbo Codes (AG-IBTCs). Furthermore, we will evaluate its performance over fast fading channels using different modulation schemes (BPSK, QPSK, 16QAM, and 64QAM). The idea behind the design of AG-IBTC is to overcome the high system complexity of Algebraic-Geometric Block Turbo Codes (AG-BTCs) while maintaining a high Bit-Error-Rate (BER) performance.

This design is inspired by the idea of unequal protection of information symbols which is the core of IBTCs. Our experiments conducted over the worst case of fading channels which is the fast fading channels; and this is done in order to show the magnitude of gain in BER or the reduction in system complexity that can be achieved by this design.

Our experiments shows that despite a very slight negative gain over BPSK and QPSK modulation schemes the system complexity is significantly reduced. However, when applying higher modulation schemes a gain in both BER and system complexity are achieved.

## INTRODUCTION

The need for low complexity systems while maintaining the BER performance is a demand of many real-time data services such as multimedia applications and deep space communications. This need has led to the introduction of Irregular Block Turbo Codes (IBTCs) in 1999 by Frey and MacKay (Frey and Mackay 1999) (Zhou et al 2004). In recent years, IBTCs draw the attention of scientists all over the world due to the fact that they have several advantages over Block Turbo Codes (BTCs). IBTCs have the following properties (Alzubi et al 2014) (Johnston and Carrasco 2005):

- Improved the BER performance compared to the BTCs.
- Reduced the decoding complexity compared to the regular (equal protection) codes.
- Produced codes closer to Shannon's bound from the BTCs.

Due to the aforementioned properties, IBTCs played an essential role in today's real time communications such as wireless communications and multimedia applications.

This paper will present the design of AG-IBTC and compare its performance with AG-BTC over fast fading channel using different modulation schemes.

**RELATED WORK**

Irregular Block Turbo Codes emerged for the first time ever in 1999 by Fery and MacKay (Frey and Mackay 1999) at the 37-th Alberton conference on communication, control and computing. At that time, they showed that by changing the structure of the Turbo Code (TC) of rate $\frac{1}{2}$ -which is introduced by Berrou *et al.* – to be slightly irregular, a coding gain of 0.15 dB is obtained at BER of $10^{-4}$ using BPSK modulation scheme over Additive White Gaussian Noise (AWGN) channel model.

In year 2000, Fery and MacKay (Frey and Mackay 2000) again presented a new paper titled "Irregular Turbo- Like Codes" in which they claimed that an ITC can be produced by increasing the number of low-weight codewords that is obtained by increasing the rate of the code components of which the irregular codes constructed from. At the decoding side of this ITC, an iterative process can be applied by using the same sum product algorithm.

However, the design presented by Ferry and MacKay (Frey and Mackay 1999) (Frey and Mackay 2000) required a large frame size which is considered as a drawback despite the fact that no information available on the number of iterations required to achieve a low BER performance.

A year later, Richardson et al. (Richardson et al 2001) presented in their paper titled "Design of Capacity-Approaching Irregular Low-Density Parity-Check Codes" a new irregular low-density parity-check (LDPC) code which has a length of 106 bits. This code performed closer to Shannon's bound than the codes presented earlier using BPSK modulation scheme over AWGN channel model. The reported performance coding gain obtained by the new code was about 0.82 dB which is 0.13 dB away from Shannon's capacity at BER of $10^{-6}$. This great improvement in performance was at the expense of higher system complexity.

H. Sawaya and J. Boutros (Sawaya and Boutros 2003) presented a new design of ITC consisting of a single Recursive Systematic Convolutional (RSC) encoder and a one Single Input Single Output (SISO) decoder. This design achieved a gain of about 0.24 dB at BER of $10^{-6}$ comparing to the regular TC using BPSK modulation scheme over AWGN channel model. Although their objective was to reduce the decoding complexity – which is an important factor in channel coding – the system required a very large frame size and suffered from high number of iterations (almost 100) in order to obtain a lower BER (e.g. $10^{-6}$).

Later, in 2011 a PhD thesis entitled "Irregular Block Turbo Codes for communication systems" written by A. Sholiyi (Sholiyi 2013) proposed a lower complexity, flexible, and fast ITC codec. The idea behind this new design is to benefits from extra protection of some bits sets in a specific manner using state of the art techniques.

In (Sholiyi 2013) the author claimed that the experimental results using this codec over AWGN gave a coding gain directly proportional to the modulation index. In others words the best gain achieved was when 64QAM modulation scheme used and of course that in comparison with existing BTCs. In summary, the aim of all codec designs in the existing literature is to improve the BER performance in order to get closer to Shannon's bound.

However, all designs suffer from high complexity which is a high price to pay (Alzubi et al 2014).

**SYSTEM DESIGN**

    **I.    Encoder Design of Irregular Algebraic-Geometric Block Turbo Codes**

In this section we will fully explain the encoder design of Irregular Algebraic-Geometric Block Turbo Codes (I-AGBTCs). In general, the principle of encoding in Block Turbo Codes (BTCs) is a simple process consists of the following steps (Soleymani et al 2002) (Johnston et al 2004) as illustrated in figure 1:

- Arranging the information bits into a block format.
- Passing the block format information bits to a systematic block encoder: In other words multiplying the information bits block by a systematic generator matrix which is constructed according a set of rules depending on the type of the code in use.
- Passing the output from previous step to a block interleaver: The interleaving process is simply interchanging the columns with the rows and vice versa.
- Passing the output from the interleaver into another systematic block encoder of the same type as the first one.

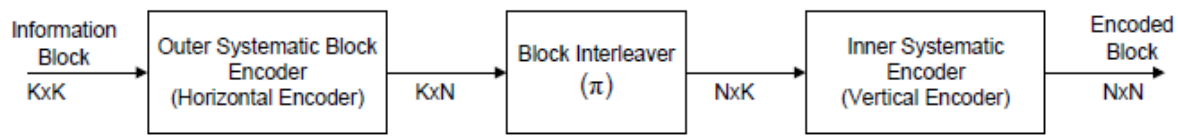

Figure 1: Conventional Block Encoding Method.

It is worth to mention here that there exists an equivalent method for designing the encoder of Turbo Product Codes (TPCs) (Pyndiah et al 1995). However, in case TPC encoders design only one encoding component used which is illustrated in figure 2.

Also when the information bits are even they are repeated (i.e. twice) because that each information bit will have one extrinsic information value from the outer decoder and another one from the inner decoder. This is similar to degree two IBTC in which each information bit in the codeword will result in two extrinsic information values in the decoding side (Sholiyi 2013).

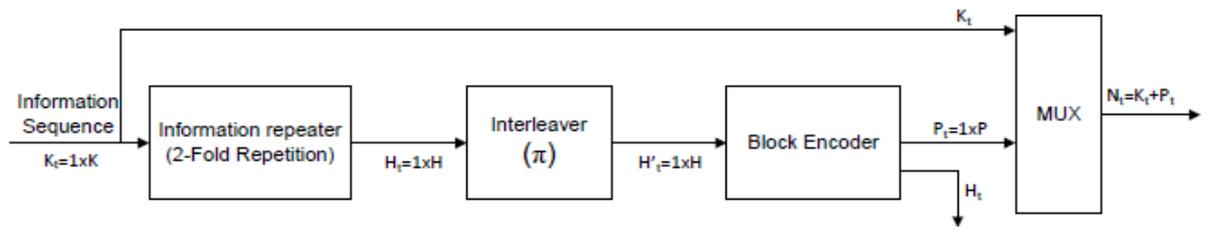

Figure 2: Equivalent Irregular Encoding Structure for TPCs.

The concept of the equivalent method explained above is applied in the AG-IBTC encoder design and construction. In AG-IBTC will assume a variable called *d* that represents the number of times that the non-binary symbols are repeated while keeping in mind the minimum value of *d* must be at least 2. However, a greater value of *d* means stronger protection on the symbols because of *a posterior* value of those symbols will be derived from *d* number of extrinsic information symbols (Xing et al 1999).

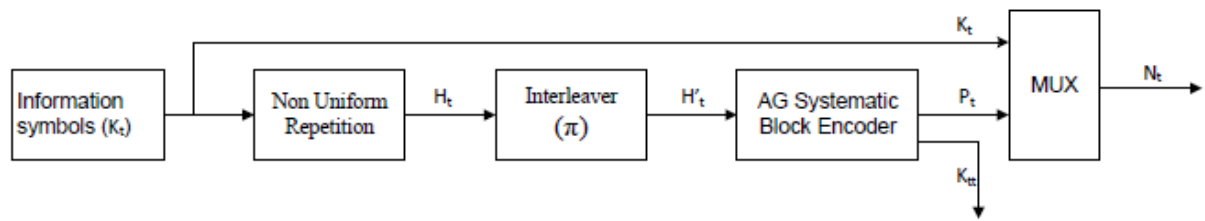

Figure 3: Encoding Structure for Irregular BTC.

A block diagram of IBTC encoder design is shown in figure 3. The encoding process starts as the information symbols $K_t$ will be passed into a non-uniform repetition unit in which the information symbols are divided into *j* groups (j = 1, 2, 3, ..., n). Where each $j^{th}$ group consists of a fraction of information symbols ($f_j$). In order to achieve a good coding performance *j* should not exceed 3. Now each $j^{th}$ group is repeated $d_i$ times, where $d_i$= 2, 3, ..., *T* . Where *T* represents the maximum repetition times. In other words, the non-uniform repetition unit will repeat $f_i$ as many as $d_i$ times.

$$A \sum_{j=1}^{3} f_i = K_t \qquad (1)$$

So, the output from the non-uniform repetition unit will be in the form of:

$$H_t = \sum_{i=2}^{T} \sum_{j=1}^{3} d_i f_i \qquad (2)$$

The choice of the right symbol degree and the appropriate corresponding fraction (denoted by $d_i$ and $f_i$ respectively) depend on the codeword size. However, there is no algorithm found to calculate the optimal values of these variables.

Usually, the fraction $f_i$ of information symbols contained in the symbol degree profile is repeated one time (degree two). The value of degree two $f_i$ preferably to be in the range of 75%-95% of the original information symbols and the remaining fraction is shared by the higher degrees depending on the modulation scheme used (Sholiyi 2013) (Pyndiah et al 1995).

Considering the aforementioned criteria in designing our AG-IBTC encoder we have chosen 85% of the information symbols to be repeated one time (degree two), 10% of the information symbols to be repeated two times (degree three), and 5% of the information symbols to be repeated eight times (degree nine) to form AG-IBTC(64, 49).

As illustrated in figure 3 the outcome from the non-uniform repetition unit is passed to the random interleaver unit, which then handled by AG systematic block encoder.

The original information symbols $K_t$ (a copy of information symbols before entering the non-uniform repetition unit) along with parity bits $P_t$ (extracted from the output of AG systematic encoder) are transmitted together in block format $N_t$.

The code rate of the obtained AG-IBTC is:

$$R = \frac{K_t}{K_t + P_t} \qquad (3)$$

Where R represents the code rate of the IBTC, and $K_t$, $P_t$ are the lengths of original information and parity respectively (Martin and Taylor 2002) (Heegard et al 1995).

For fairer BER performance comparisons, the block size of the corresponding regular BTC limits and controls $d_i$ which plays an important role in the construction of the IBTC.

## II. Decoder Design of Irregular Algebraic-Geometric Block Turbo Codes

The received sequence $N_r$ in handled by a proper demodulator in order to be able to pass it through demultiplexer which separates the parity symbols $P_r$ from information symbols $K_r$ for each encoded message $N_r$.

Using a similar non-uniform repetition unit as the one used in the transmitter side the information symbols part $K_r$ is repeated in the same manner to obtain a repeated information block $H_r$ which passed to a similar random interleaver as the one in the transmitter side again (Frey and Mackay 1999) (Sholiyi 2013).

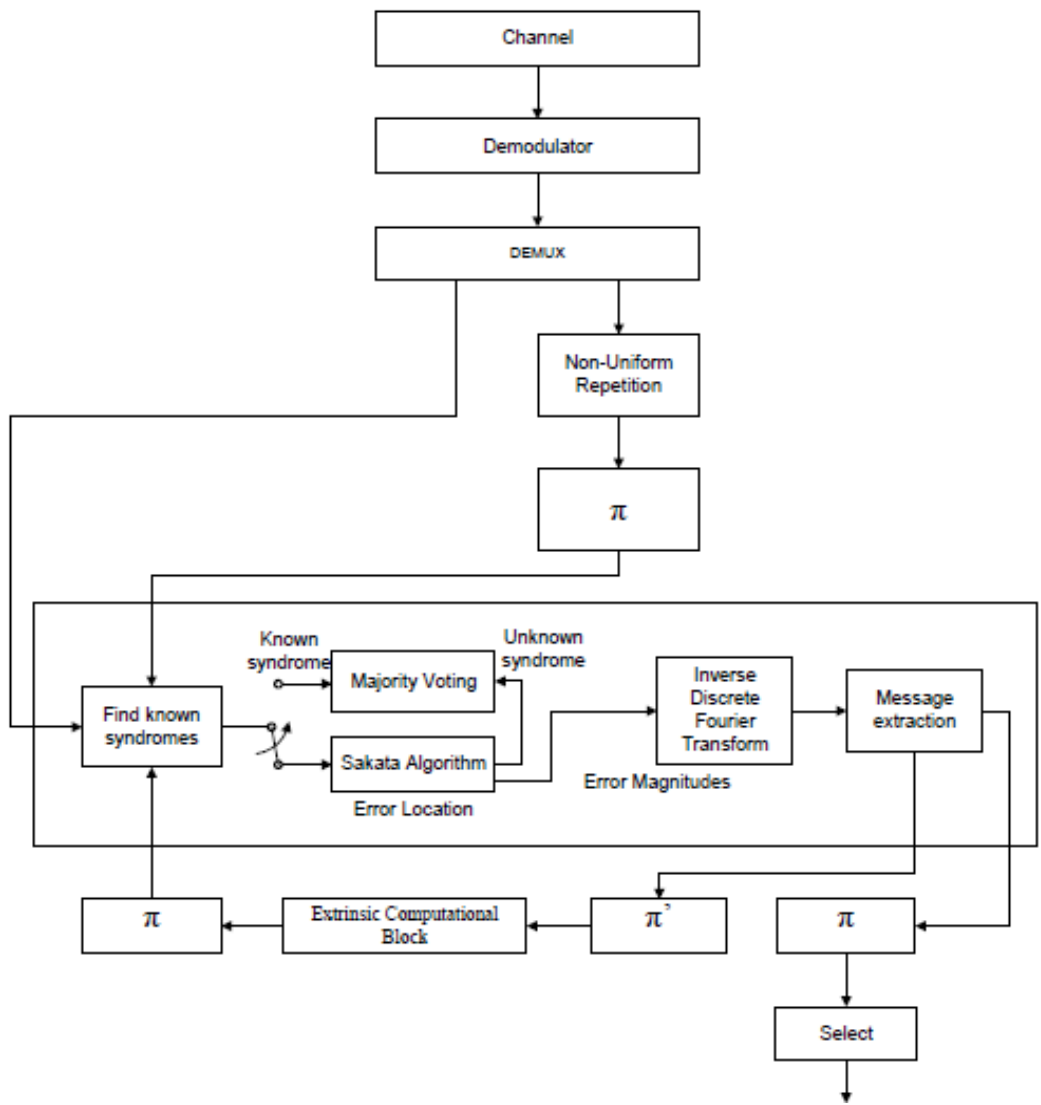

Figure 4: Decoder Structure for Irregular BTC.

The interleaved vector of repeated information $h'_r$ is combined with its corresponding parity vector $P_r$ to form a new vector $n_r$. Before passing $n_r$ to the AG decoder an initial a priori value $a_r$ of equal probability (i.e. zero log likelihood) is added.

From the output of the AG decoder an extrinsic information $e_r$ - of same size as vector $h'_r$ - is gathered and computed, a set of $e_r$ vectors is arranged in block form $E_r$ then passed to a random deinterleaver before entering the extrinsic computational block (Frey and Mackay 1999). In every iteration new extrinsic information value is calculated for every information symbol of degree $d_i$. The new extrinsic information found by multiplying $d_i-1$ extrinsic information values by the sum of those values when using log likelihood values (Richardson et al 2001) (Skorobogatov and Vladut 1990) (Berrou and Glavieux 1996).

By randomly interleaving the output of the extrinsic of the computational block the new a priori values block $A_r$ is computed.

To be able to add the a priori vectors ai to the nr vectors for the next decoding iteration, each $a_i$ will be read out individually from $A_i$.

A block of decoded codewords is obtained by storing each decoded codeword after the final iteration. By using a random deinterleaver, the parity part of each codeword is removed in order to extract only the information part to retain the original information vector $K_t$ format (Alzubi et al 2014) (Sholiyi 2013) (Pyndiah 1998).

Figure 4, clearly shows the explained decoding process. It is important to mention here that the random interleaver and random deinterleaver patterns are totally different for each data block.

**RESULTS AND DISCUSSION**

The proposed design of AG-IBTC reduced the system complexity due to the fact that only one encoding and one decoding components have been used in comparison with two of each in the AG-BTC.

Although the proposed design of AG-IBTC required more iterations for the same BER performance the system complexity still considered less than the one in AG-BTC. Our objective here is to explore the trade-off between the system complexity and performance and emphasize on the conditions where AG-IBTC outperforms the AG-BTC.

Our experiments have been conducted over Rayleigh fading channel using different modulation schemes.

The figures we presented here are intentionally did not show the BER performance for every iteration in order to keep them neat and easy to visualize. In order to conduct a fair comparison between AG-IBTC and AG-BTC we used similar data block sizes and almost same code rates over the same finite field $GF(2^4)$.

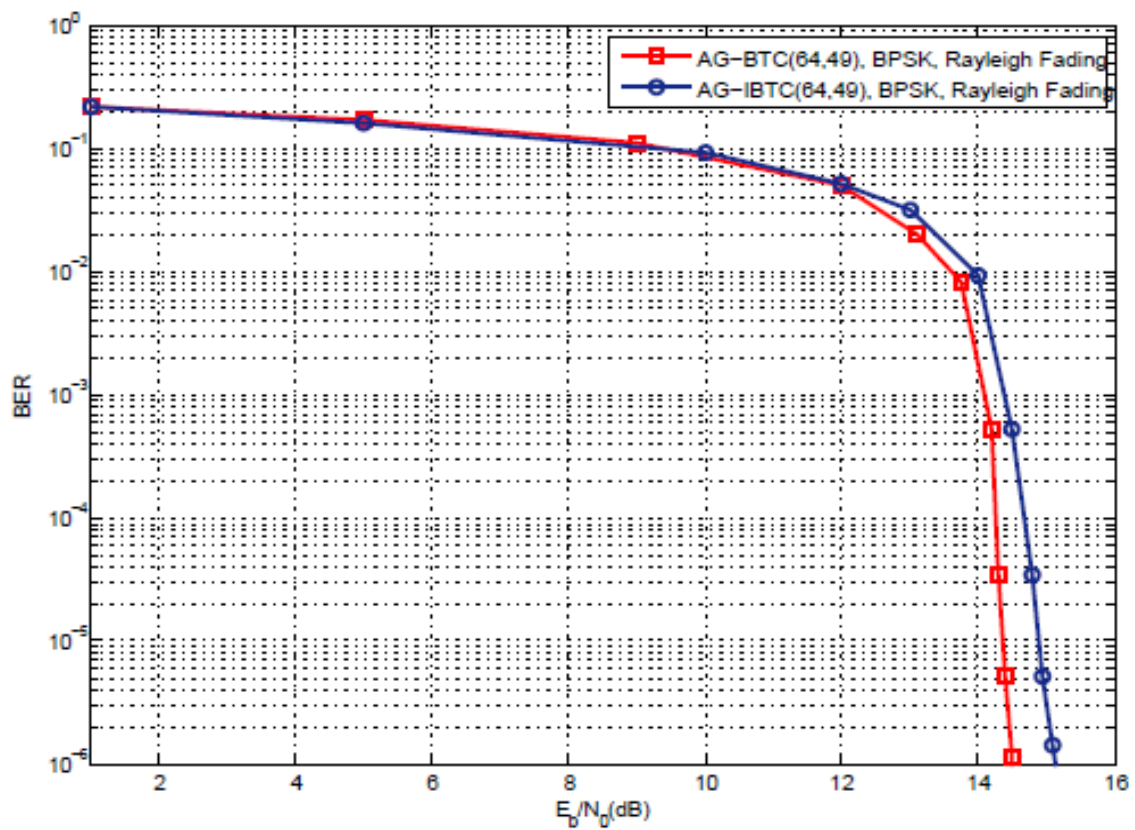

Figure 5: BER of AG-IBTC(64,49) vs. AG-BTC(64,49) using BPSK over Rayleigh fading channel.

Using BPSK modulation scheme over Rayleigh fast fading channel as shown in Figure 5 and Figure 6, the coding gains in terms of BER performance of AG-IBTC codes at BER of $10^{-6}$ are

-0.6 and -0.5 *dBs* with code rates of 0.57 and 0.5 respectively in comparison to AG-BTC codes of code rates 0.585 and 0.47. The losses from using the AG-IBTCs design seems considerable from the first glance. However, note these losses are applicable to high $E_B/N_o$ region and come as the result of severe fading channel conditions. Moreover, AG-IBTC allows us to enjoy huge system complexity reduction which is highly desirable in severe fading conditions.

Figure 7 and Figure 8 show the QPSK results over Rayleigh fast fading channel. The coding gains in terms of BER performance of AG-IBTC codes at BER of $10^{-6}$ are -0.5 and -0.4 *dBs* with code rates of 0.57 and 0.5 respectively in comparison to AG-BTCs of code rates 0.585 and 0.47. Although the losses are again negligible, they are decreasing as the modulation index increases to QPSK.

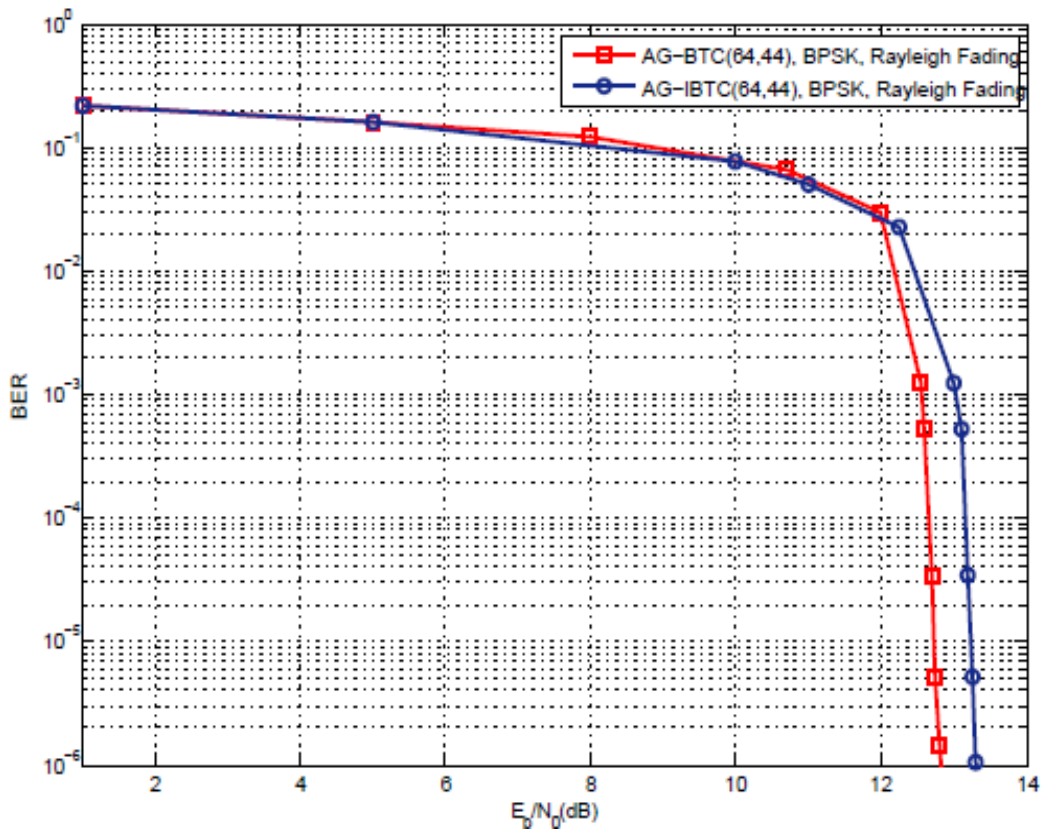

Figure 6: BER of AG-IBTC(64,44) vs. AG-BTC(64,44) using BPSK over Rayleigh fading channel.

Figure 9 and Figure 10 show the 16QAM results over Rayleigh fast fading channel. The coding gains in terms of BER performance of AG-IBTC codes at BER of $10^{-6}$ are 0.6 and 0.65 *dBs* with code rates of 0.57 and 0.5 respectively in comparison to AG-BTCs of code rates 0.585 and 0.47. Again note here that the gains are positive and significant. This re-assures us about the versatility of AG codes when used in IBTC design.

Figure 11 and Figure 12 show the 64QAM results over Rayleigh fast fading channel. The coding gains in terms of BER performance of AG-IBTC codes at BER of $10^{-6}$ are 0.8 and 0.8 *dBs* with code rates of 0.57 and 0.5 respectively in comparison to AG-BTCs of code rates 0.585 and 0.47. Again the maximum coding gains are achieved in the 64QAM modulation confirms our previous conclusions and providing an interesting performance-complexity trade-off.

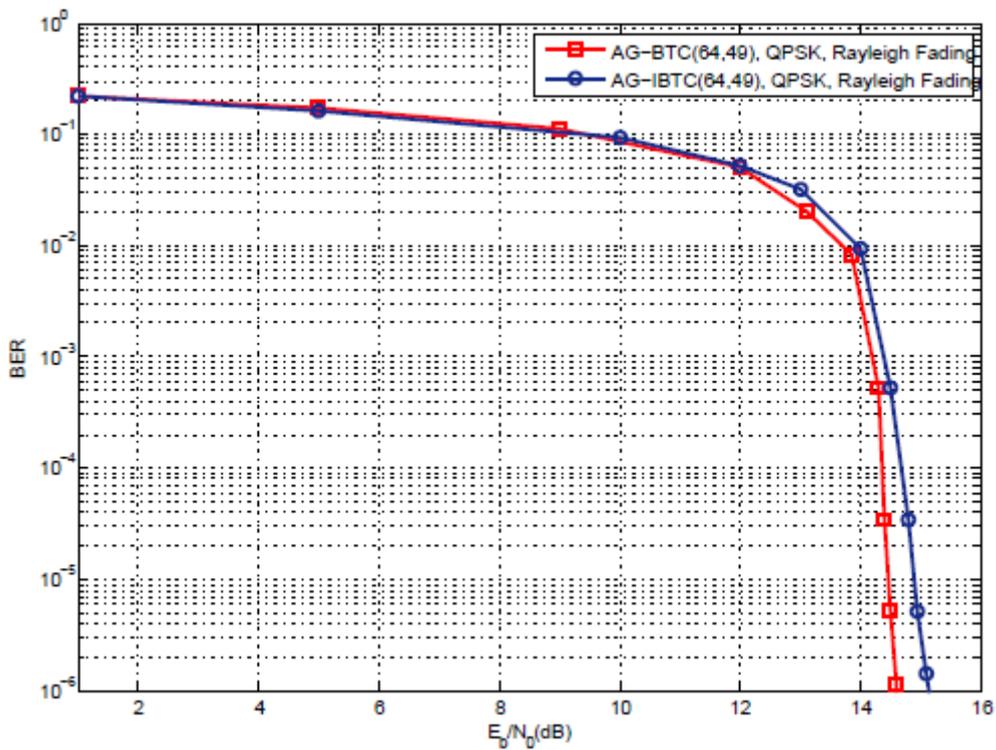

Figure 7: BER of AG-IBTC(64,49) vs. AG-BTC(64,49) using QPSK over Rayleigh fading channel.

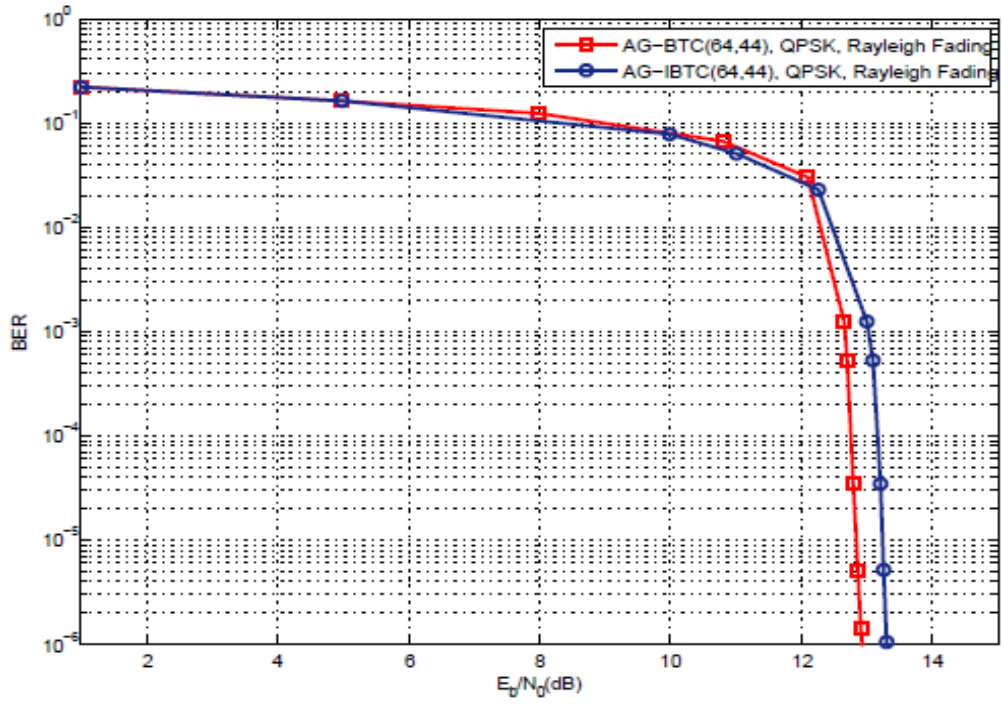

Figure 8: BER of AG-IBTC(64,44) vs. AG-BTC(64,44) using QPSK over Rayleigh fading channel.

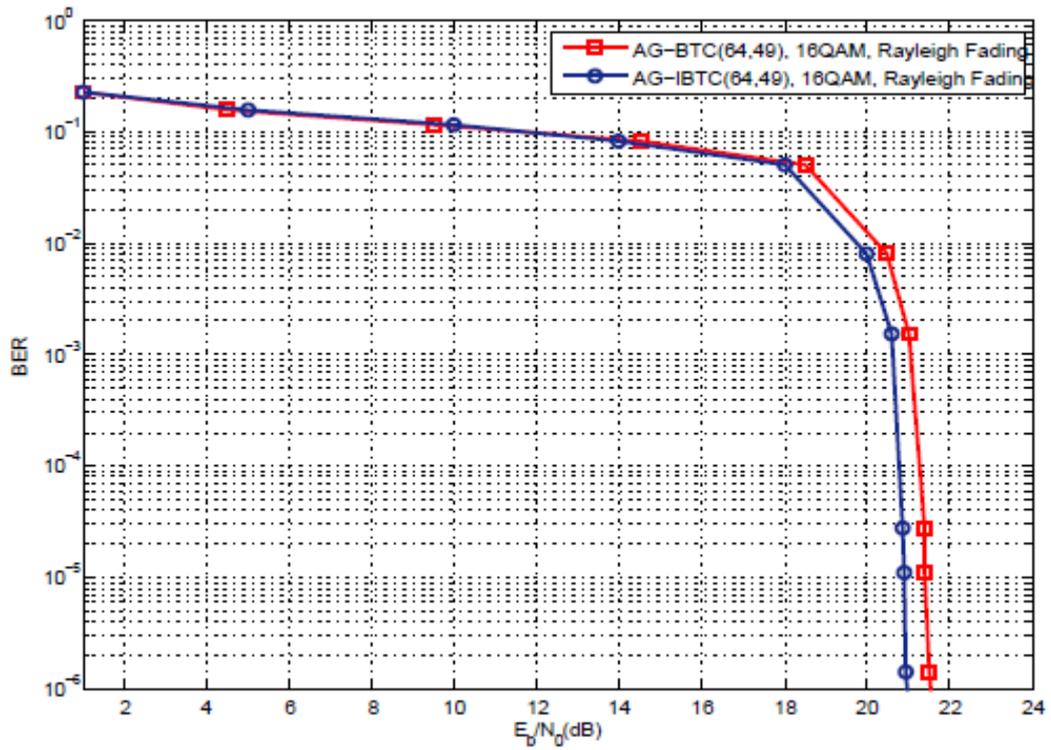

Figure 9: BER of AG-IBTC(64,49) vs. AG-BTC(64,49) using 16QAM over Rayleigh fading channel.

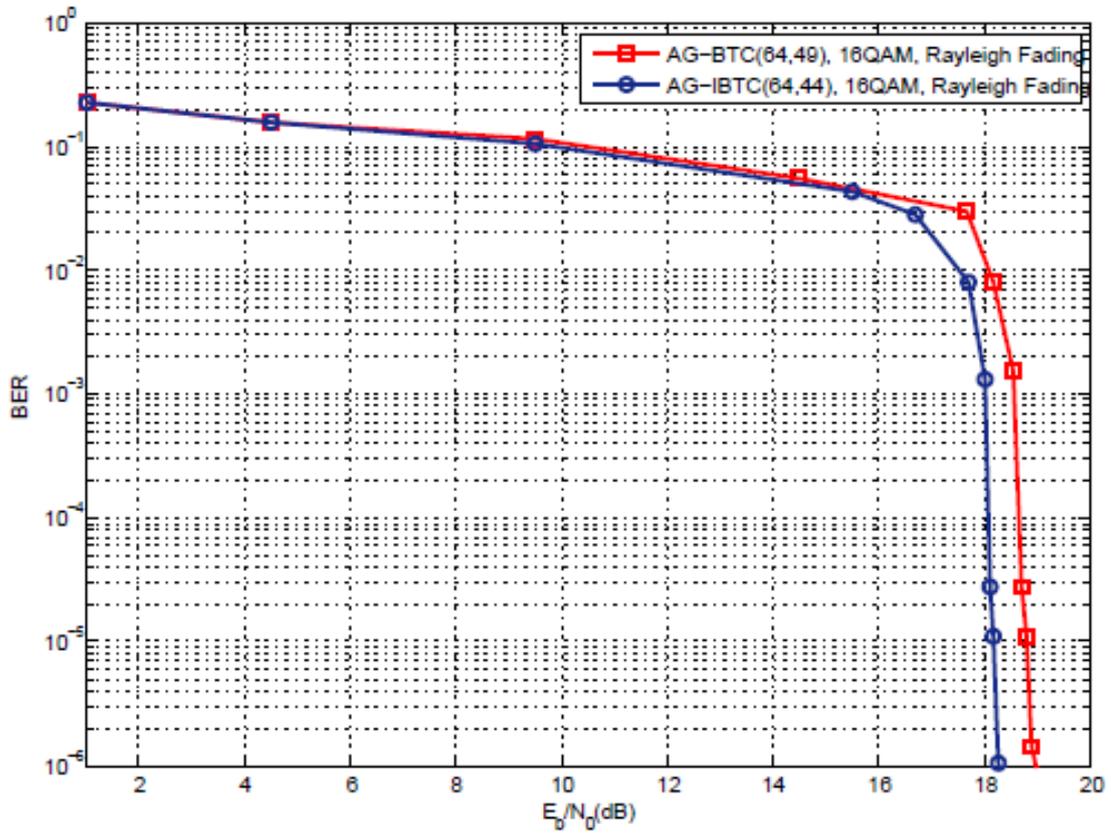

Figure 10: BER of AG-IBTC(64,44) vs. AG-BTC(64,44) using 16QAM over Rayleigh fading channel.

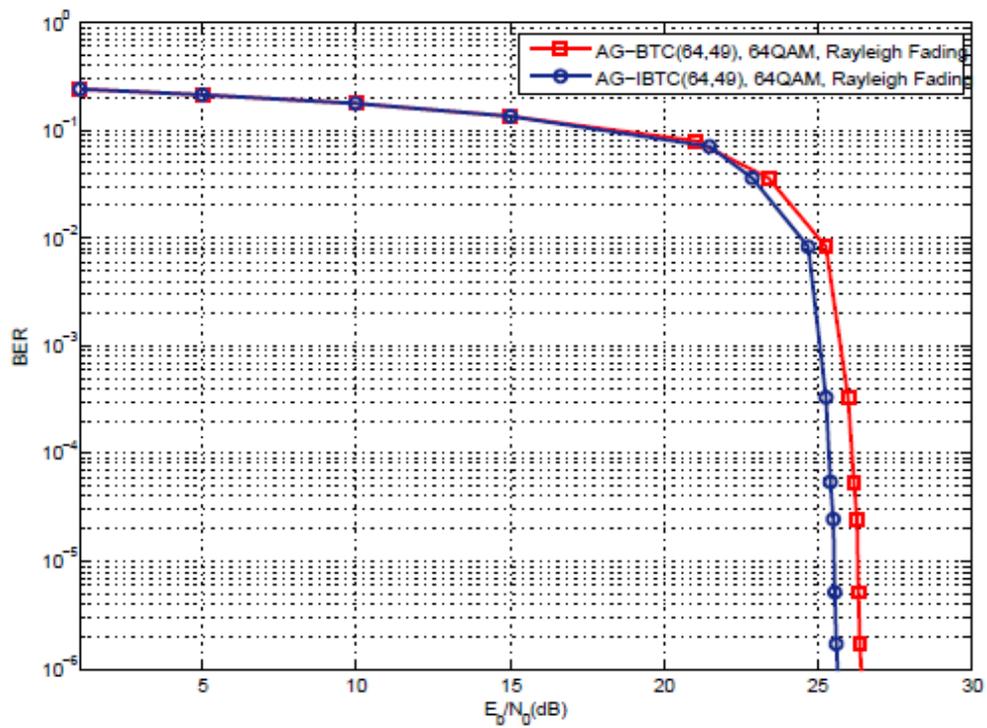

Figure 11: BER of AG-IBTC(64,49) vs. AG-BTC(64,49) using 64QAM over Rayleigh fading channel.

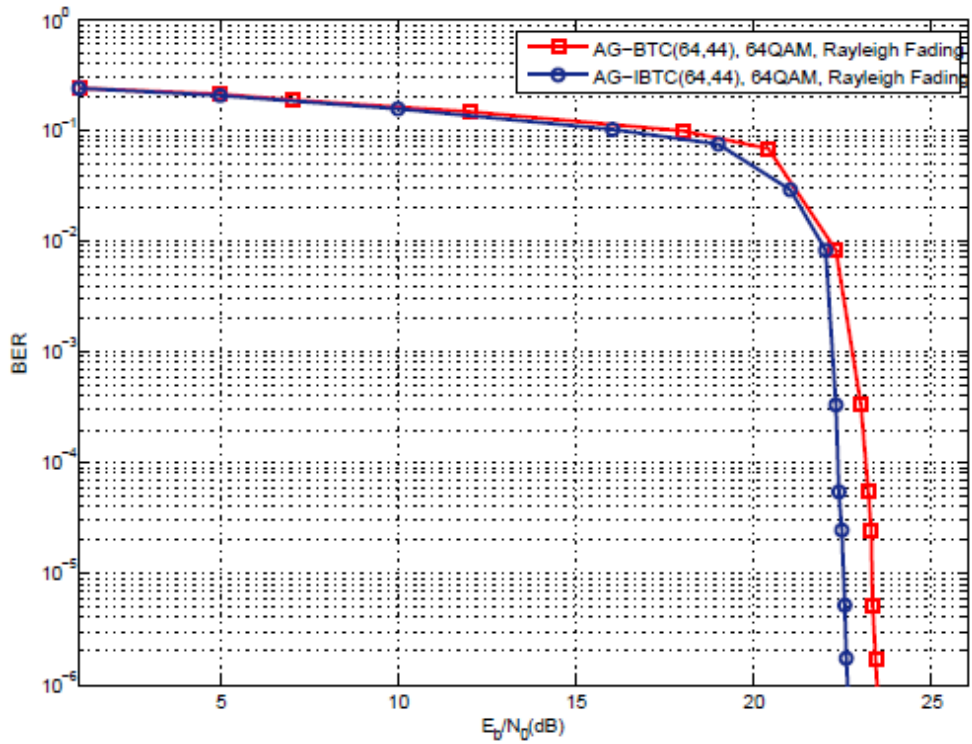

Figure 12: BER of AG-IBTC(64,44) vs. AG-BTC(64,44) using 64QAM over Rayleigh fading channel.

**CONCLUSION**

In this paper, our aim was to reduce the system complexity while keeping in mind to maintain the BER performance. This was achieved by constructing a new AG-IBTC.

Empirical results obtained using the developed MATLAB simulator to measure the BER performance of AG-IBTCs compared to their equivalent AG-BTCs over Rayleigh fast fading channel model. The comparison is performed on similar data block length, code rates and over the same finite field.

For the Rayleigh fast fading channel model, AG-IBTC results in 0.6 and 0.5 *dBs* coding loss at BER of $10^{-6}$ for code rates 0.57 and 0.5 respectively using BPSK modulation. QPSK modulation results in a slightly reduced coding loss. For 16QAM and 64QAM the coding gains become positive i.e. 0.6 and 0.65 dBs and 0.8 and 0.8 dBs for code rates 0.57 and 0.5 respectively at BER of $10^{-6}$.

Such transition from negative to positive confirms the fact that the AG codes in general and specifically AG-IBTCs gain increase with modulation index increase. Also it gives a solution to the complexity issue of AG-BTC.

We conclude here that the AG code is much resilient to fading even if it is used in BTC and IBTC. In AG-IBTCs this comes with additional benefit of significant reduction in system's complexity.

**ACKNOWLEDGMENT**

The author of this paper would like to deeply thank Dr. Jafar A. Alzubi at Al-Balqa Applied University for his invaluable advices.

**REFERENCES**

[1] Alzubi J., Alzubi O., and Chen T., 2014, "Forward Error Correction Based On Algebraic-Geometric Theory", Springer, ser. Springer Briefs in Electrical and Computer Engineering.

[2] Berrou C., and Glavieux A., 1996, "Near optimum error correcting coding and decoding: Turbo-codes," Communications, IEEE Transactions on, vol. 44, pp. 1261-1271.

[3] Frey B., and Mackay D., 1999, "Irregular turbo codes," in proceedings of the 37th Allerton conference on communication, control and computing, (Allerton House, Illinois).

[4] Frey B., and Mackay D., 2000, "Irregular turbo-like codes," In proceedings of the 2nd International Symposium on Turbo Codes and related topics, (Brest, France), pp. 67-72.

[5] Heegard C., Little J., and Saints K., 1995, "Systematic encoding via grobner bases for a class of algebraic-geometric goppa codes," Information Theory, IEEE Transactions on, vol. 41, pp. 1752-1761.

[6] Johnston M., Carrasco R., and Burrows B., 2004, "Design of algebraic-geometric codes over fading channels," Electronics Letters, vol. 40, pp. 1355-1356.

[7] Johnston M. and Carrasco R., 2005, "Construction and performance of algebraic-geometric codes over AWGN and fading channels", Communications, IEE Proceedings-, vol. 152, pp. 713-722.

[8] Martin P., and Taylor D., 2002, "Distance based adaptive scaling in suboptimal iterative decoding," Communications, IEEE Transactions on, vol. 50, pp. 869-871.

[9] Pyndiah R., Picart A., and Glavieux A., 1995, "Performance of block turbo coded 16-qam and 64-qam modulations," in Global Telecommunications Conference, 1995. GLOBECOM '95., IEEE, vol. 2, pp. 1039-1043.

[10] Pyndiah R., 1998, "Near-optimum decoding of product codes: Block turbo codes," Communications, IEEE Transactions on, vol. 46, pp. 1003-1010.

[11] Richardson T., Shokrollahi M., and Urbanke R., 2001, "Design of capacity- approaching irregular low-density parity-check codes," Information Theory, IEEE Transactions on, vol. 47, pp. 619-637.

[12] Sawaya H. E., and Boutros J. J., 2003, "Irregular turbo codes with symbol-based iterative decoding," In proceedings of the 3rd International Symposium on Turbo Codes and related topics, (Brest, France).

[13] Sholiyi A., 2011, "Irregular Block Turbo Codes For Communication Systems," PhD thesis, Swansea University, United Kingdom.

[14] Skorobogatov A., and Vladut S., 1990, "On the decoding of algebraic-geometric codes," Information Theory, IEEE Transactions on, vol. 36, pp. 1051-1060.

[15] Soleymani M., Gao Y., and Vilaipornsawai U., 2002, "Turbo Coding For Satellite And Wireless Communications". Vol. V. No. 1 in Kluwer international series in engineering and computer science, Kluwer Academic Publishers.

[16] Xing C., Niederreiter H., and Lam K. Y., 1999, "Constructions of algebraic-geometry codes," Information Theory, IEEE Transactions on, vol. 45, pp. 1186-1193.

[17] Zhou R., Picart A., Pyndiah R., and Goalic A., 2004, "Potential applications of low complexity non-binary high code rate block turbo codes," in Military Communications Conference, 2004. MILCOM 2004. IEEE, vol. 3, pp. 1694-1699.